\documentclass[sn-mathphys,Numbered]{sn-jnl}% Math and Physical Sciences Reference Style
\usepackage{graphicx}%
\usepackage{multirow}%
\usepackage{amsmath,amssymb,amsfonts}%
\usepackage{amsthm}%
\usepackage{mathrsfs}%
\usepackage[title]{appendix}%
\usepackage{xcolor}%
\usepackage{textcomp}%
\usepackage{manyfoot}%
\usepackage{booktabs}%
\usepackage{algorithm}%
\usepackage{algorithmicx}%
\usepackage{algpseudocode}%
\usepackage{listings}%
\usepackage{bm}
\theoremstyle{thmstyleone}%
\theoremstyle{thmstyletwo}%

\theoremstyle{thmstylethree}%

\begin{document}

\title[Article Title]{Specialty-Oriented Generalist Medical AI for Chest CT Screening}


\author[1]{\fnm{Chuang} \sur{Niu}}\email{niuc@rpi.edu}

\author[2]{\fnm{Qing} \sur{Lyu}}\email{qlyu@wakehealth.edu}

\author[1]{\fnm{Christopher D.} \sur{Carothers}}\email{chrisc@cs.rpi.edu}

\author[3]{\fnm{Parisa} \sur{Kaviani}}\email{pkaviani@mgh.harvard.edu}

\author[2]{\fnm{Josh} \sur{Tan}}\email{jtan@wakehealth.edu}

\author[1]{\fnm{Pingkun} \sur{Yan}}\email{yanp2@rpi.edu}

\author*[3]{\fnm{Mannudeep K.} \sur{Kalra}}\email{mkalra@mgh.harvard.edu}

\author*[2]{\fnm{Christopher T.} \sur{Whitlow}}\email{cwhitlow@wakehealth.edu}

\author*[1]{\fnm{Ge} \sur{Wang}}\email{wangg6@rpi.edu}

\affil[1]{\orgdiv{Department of Biomedical Engineering,
School of Engineering,
Center for Computational Innovations,
Center for Biotechnology \& Interdisciplinary Studies}, \orgname{Rensselaer Polytechnic Institute}, \orgaddress{\street{110 8th Street}, \city{Troy}, \postcode{12180}, \state{NY}, \country{USA}}}

\affil[2]{\orgdiv{Department of Radiology}, \orgname{Wake Forest University School of Medicine}, \orgaddress{\city{Winston-Salem}, \postcode{27103}, \state{NC}, \country{USA}}}

\affil[3]{\orgdiv{Department of Radiology}, \orgname{Massachusetts General Hospital, Harvard Medical School}, \orgaddress{\street{White 270-E, 55 Fruit Street}, \city{Boston}, \postcode{02114}, \state{MA}, \country{USA}}}


\abstract{
Modern medical records include a vast amount of multimodal free text clinical data and imaging data from radiology, cardiology, and digital pathology. Fully mining such big data requires multitasking; otherwise, occult but important aspects may be overlooked, adversely affecting clinical management and population healthcare. Despite remarkable successes of AI in individual tasks with single-modal data, the progress in developing generalist medical AI (GMAI) remains relatively slow to combine multimodal data for multitasks because of the dual challenges of data curation and model architecture. The data challenge involves querying and curating multimodal structured and unstructured text, alphanumeric, and especially 3D tomographic scans on an individual patient level for real-time decisions and on a scale to estimate population health statistics. The model challenge demands a scalable and adaptable network architecture to integrate multimodal datasets for diverse clinical tasks. Here we propose the first-of-its-kind medical multimodal-multitask foundation model (M3FM) with application in low-dose CT lung cancer screening and related tasks. After we curated a comprehensive multimodal multitask dataset consisting of 49 clinical data types including 163,725 chest CT series and 17 medical tasks involved in lung cancer screening, we develop a multimodal question-answering framework as a unified training and inference strategy to synergize multimodal information and perform multiple tasks via free-text prompting. Our M3FM consistently and significantly outperforms the state-of-the-art single-modal task-specific models, improving lung cancer risk prediction by up to 20\% and cardiovascular disease mortality risk estimation by up to 10\%. M3FM identifies multimodal data elements informative for clinical tasks, being instrumental in gaining insights and correlating the multimodal data with relevant diseases. M3FM can flexibly adapt to new tasks with a small out-of-distribution dataset, effectively handle various combinations of multimodal data, and efficiently process high-dimensional images at multiple scales. In a broader sense, as a specialty-oriented generalist medical AI (SOGMAI) model, M3FM innovates lung cancer management and related tasks. This SOGMAI approach paves the way for similar breakthroughs in other areas of medicine, closing the gap between specialists and the generalist.
}

\maketitle

\section{Introduction}\label{sec1}

Medical diagnosis, treatment, and management require simultaneous multiple tasks on multimodal data, such as unstructured medical records (for symptoms, comorbidities, risk factors, etc.), structured text data (such as laboratory test results), and imaging data (such as computed tomography (CT) images). In the fast-evolving domain of artificial intelligence (AI), foundation models (FMs) have shown previously unseen abilities to understand diverse data types and execute many tasks in a unified architecture \cite{bommasani2021opportunities}. Large FMs have updated the state-of-the-art (SoTA) performance across a wide range of tasks, such as natural language processing \cite{brown2020language,anil2023palm,touvron2023llama2}, computer vision \cite{dehghani2023scaling, kirillov2023segment}, and vision-language understanding \cite{radford2021learning, chen2022pali}.
Propelled by the breakthroughs of FMs, the medical community calls for generalist medical AI (GMAI) \cite{moor2023foundation}. GMAI models are expected to perform diverse tasks per instructions in free text, making predictions with different combinations of data modalities and interpreting the outputs as medical professionals.

Along this direction, biomedical language models \cite{li2023chatdoctor,han2023medalpaca,wu2023pmc,toma2023clinical,xiong2023doctorglm,wang2023huatuo} were developed by fine-tuning open-source \cite{touvron2023llama1, touvron2023llama2} or closed-source \cite{chowdhery2022palm,chung2022scaling} large language models (LLMs) on the customized biomedical datasets.
Among these models, Med-PaLM \cite{singhal2023large} achieves SoTA performance on various datasets due to its large model size, which demonstrates the power of scaling laws \cite{kaplan2020scaling} in the medical domain although they can only process textual data.
Most recently, Google Research and Deep Mind prepared a multimodal biomedical benchmark called MultiMedBench and developed the Med-PaLM Multimodal (Med-PaLM M) system \cite{tu2024towards}. On their test datasets, Med-PaLM M reached performance comparable to or better than the SOTA records. While it represents a milestone proof of concept, Med-PaLM M inherits the vision encoder pre-trained with natural images, only takes 2D images of limited size, handles each task mainly by taking a single modality data without systematic alignment among multiple data modalities, and accumulates relatively disconnected public datasets and medical tasks that are not well aligned with specific clinical scenarios.
These facts limit its utility in real-world clinical applications.

Beyond the above proof-of-concept studies, we believe that the efforts toward GMAI should involve the development of specialty-oriented generalist medical AI (SOGMAI) models. Integration of these SOGMAI models will bring us to the holy grail of GMAI. It is underlined that SOGMAI is both general and specific.
It is general in scope, as it encompasses a multitude of interconnected tasks and interprets a variety of correlated data modalities within each medical specialty. At the same time, the SOGMAI model is specific in alignment with a medical specialty, making it seamlessly applicable to streamline a specific clinical workflow.
Here we introduce our Medical Multimodal Multitask Foundation Model (M3FM) as the first SOGMAI model, expressly designed for lung cancer screening (LCS).
LCS involves multiple tasks, such as lung nodule detection and characterization, lung cancer risk evaluation, diagnosis of a set of chest abnormalities, COVID-19 detection, and diagnosis and risk evaluation of cardiovascular diseases (CVD); and multimodal data, including low-dose computed tomography (LDCT) images, patient demographics, smoking history, disease history, family cancer history, pathological results, follow-up data, etc \cite{national2011reduced}.

Indeed, lung cancer remains the leading cause of cancer-related deaths~\cite{lung}. Treatment at an earlier stage is much more effective and less invasive, reducing the medical cost and social burden. For LCS, 3D LDCT is the method of choice over 2D chest radiography, as LDCT screening reduced lung cancer mortality by 20\% in comparison with chest radiography in the National Lung Screening Trial (NLST) \cite{national2011reduced} and 24\% mortality reduction in comparison with no screening in the NELSON trial \cite{de2020reduced}.  
However, the LCS fraction of eligible smokers remains low relative to other screening tests (such as for cervical, colorectal, and breast cancers) due to various barriers to implementing LCS, such as resource inaccessibility \cite{wang2019barriers, fedewa2021state}, the challenging patient management \cite{tseng2019relationship, rivera2016lung, triplette2022patient, lin2022patient, nunez2021adherence}, and particularly, a global shortage of radiologists with the requisite training for providing LCS.
Hence, there is an important and immediate need for multidisciplinary efforts to broadly, equitably, and optimally implement LCS for reduced lung cancer mortality~\cite{wang2019barriers}.

Over the past years, various task-specific AI methods have been studied on the LCS datasets. For example, a deep learning method was proposed for lung cancer detection and risk estimation with LDCT in an end-to-end manner \cite{ardila2019end}.
Recently, the Sybil model \cite{mikhael2023sybil} was developed for lung cancer risk prediction using a single LDCT scan.
In some studies \cite{ruparel2019evaluation, chao2021deep}, deep learning models were developed for cardiovascular diseases (CVD) risk prediction with LDCT from LCS.
All these models are focused on narrow tasks using single modality data, limiting their performance and utility in LCS.
Due to the high dimensionality of volumetric CT images, these efforts only studied small 2D/3D ResNet models \cite{he2016deep} with affordable computation costs.
In particular, the training schemes of current lung cancer risk models \cite{ardila2019end, mikhael2023sybil} require costly bounding box annotations, which makes building large-scale training datasets impractical.

It is well recognized that given the sheer size of multimodal datasets covering diverse medical tasks, the large FMs could yield unprecedentedly better medical AI outcomes~\cite{kaplan2020scaling}, thereby revolutionizing numerous clinical scenarios.
However, progress remains relatively slow in this direction for two main reasons.
First, we face the data challenge.
There is a high bar to curate medical multimodal multitask datasets, systematically aligning 3D medical images and other diverse structured and unstructured text-based clinical data with various medical tasks in specific clinical scenarios.
For example, current data collection and curation strategies \cite{tu2023towards, li2023llavamed} are not well aligned with specific clinical scenarios.
Second, we have the model challenge.
There is no scalable and adaptable foundation model dedicated to effectively interpreting medical multimodal data, especially high-dimensional medical images at different scales, and flexibly performing diverse clinical tasks.
Although some efforts \cite{tu2023towards,li2023llavamed} have explored fine-tuning general domain foundation models for medical tasks, how to effectively encode and interpret multimodal data and how to synergistically unify diverse medical tasks remain an open challenge.

\begin{figure}
    \centering
    \includegraphics[width=1\textwidth]{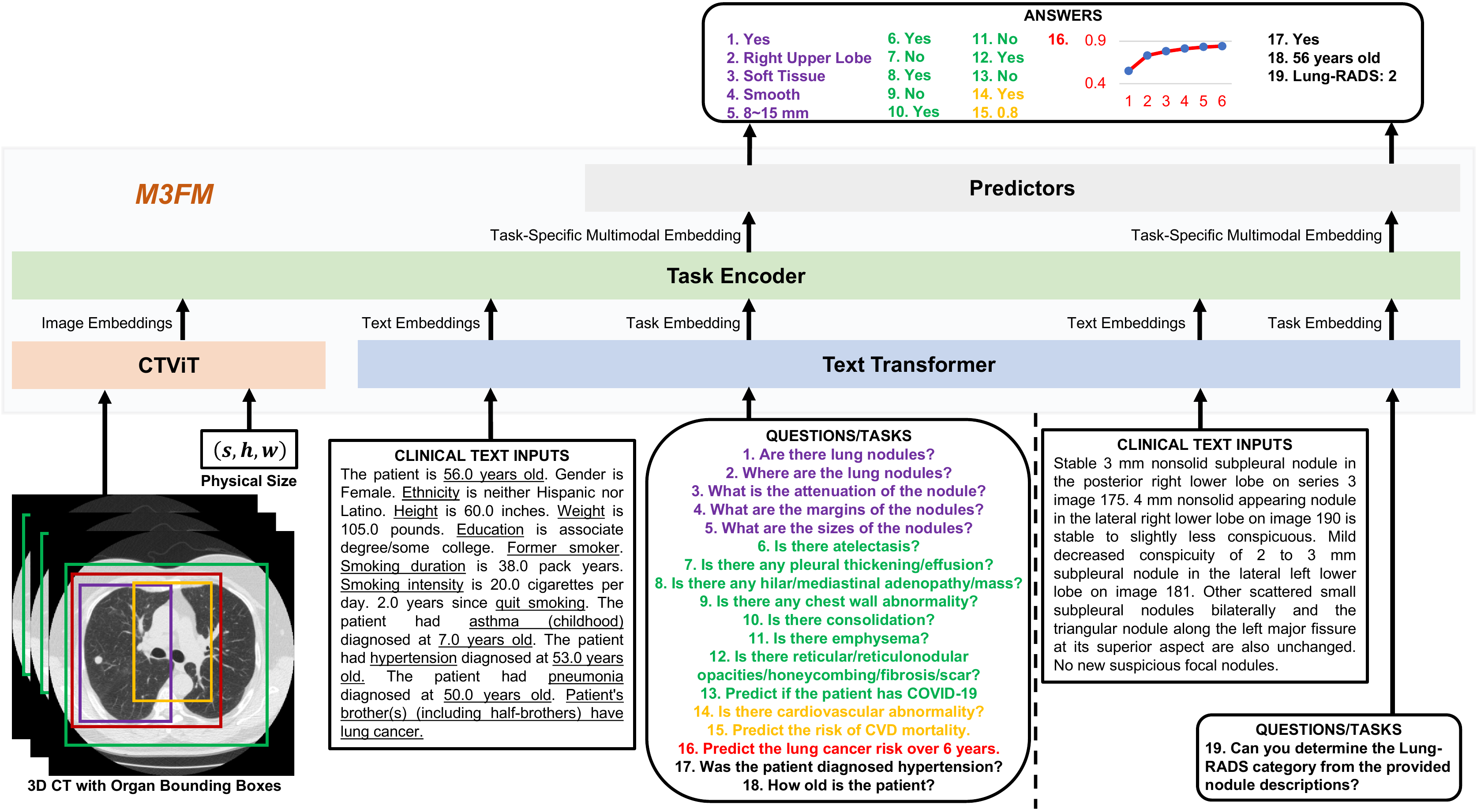}
    \caption{\textbf{M3FM inference illustration.} M3FM consists of four main components: CTViT, Text Transformer, Task Encoder, and Predictors. CTViT encodes volumetric CT images at different scales (indicated by rectangle boxes in different sizes and colors). Text Transformer encodes a combination of clinical data and questions in free text. All image and text tokens are forwarded to the Task Encoder, which extracts task-specific embedding features of the integrated multimodal data. The task-specific Predictors take task-specific embedding features to derive the final answers. These tasks may have arbitrary combinations of inputs including varied sizes of imaging data and shared/different predictors. Different colors in CT, QUESTIONS/TASKS, and ANSWERS differentiate the matches among them. The questions in the black text take the textural inputs only. Questions 17 and 18 are two examples of simulated clinical information retrieval tasks to inspect the ability of clinical data modeling. Question 19 only takes its neighboring clinical text as input.}
    \label{fig:example}
\end{figure}

In this study, we report the first-of-its-kind SOGMAI model, M3FM, that can perceive multimodal data to perform multiple tasks dedicated to LCS, as illustrated in Figure~\ref{fig:example}.
The success of our M3FM for LCS is achieved by addressing both the data and model challenges.

\begin{figure}
    \centering
    \includegraphics[width=1\textwidth]{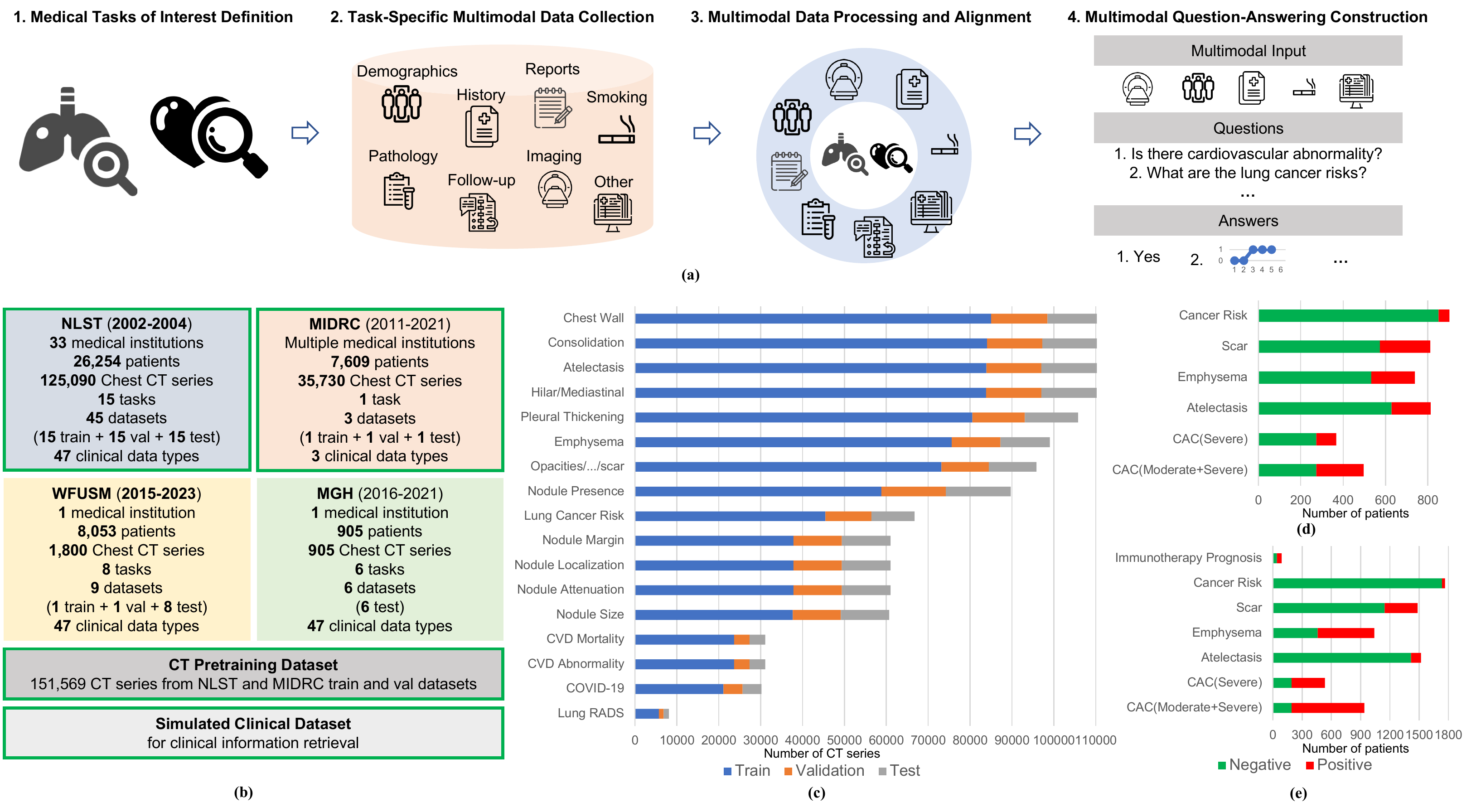}
    \caption{\textbf{Dataset construction.} (\textbf{a}) The general data construction workflow consists of four steps: medical tasks of interest definition, task-specific multimodal data collection, multimodal data processing and alignment, and multimodal question-answering construction; (\textbf{b}) The data used in this study were collected from two data centers, i.e., NLST and MIDRC, and two medical institutes, i.e., WFUSM and MGH, with the key summarized characteristics, based on which a large 3D CT pretraining dataset and a simulated clinical dataset were constructed. The green boxes indicate the OpenM3Chest dataset that will be made publicly available; (\textbf{c}) Distributions of main training, validation, and test datasets over all tasks are summarized; (\textbf{d}) Distributions of MGH independent evaluation datasets are summarized; (\textbf{e}) Distributions of WFUSM independent evaluation and fine-tuning datasets are summarized.}
    \label{fig:data_all}
\end{figure}

First, we present an integrated and scalable data curation approach to align high-dimensional medical images with other clinical datasets for specialty-oriented tasks by leveraging domain-specific expertise.
According to the LCS practice, 17 tasks of interest were defined and 49 different data elements were involved.
To collect the multimodal data for these tasks, we were granted access to massive data from the National Lung Cancer Screening Trial (NLST)~\cite{national2011reduced} and the Medical Imaging and Data Resource Center (MIDRC)~\cite{midrc}, both of which involved multiple medical institutes. Also, we independently collected two similar datasets from Wake Forest University School of Medicine (WFUSM) and Massachusetts General Hospital (MGH).
Our datasets were curated by aligning among a variety of data types for each task, including chest CT scans and the associated imaging parameters, radiology reports, de-identified demographics, smoking history, disease history, cancer history, family history, pathological results, and follow-up data.
The resulting 64 sub-datasets for training, validating, testing, and fine-tuning are summarized in Figure~\ref{fig:data_all} and Tabel~\ref{table_mqa}.
It is worth noting that we used only medical records without expensive human labeling, such as bounding box annotations for lung nodules or other imaging findings, during the data curation so that these datasets could be easily enlarged.

Second, we develop a scalable and adaptable M3FM architecture with an emphasis on LCS with LDCT. M3FM can not only analyze different combinations of medical multimodal data including multi-scale 3D tomographic scans but also perform multiple tasks in a unified manner, the architecture details are shown in Figure~\ref{fig:method}.
Routinely, radiologists undertake multiple tasks, drawing upon multimodal information to generate a comprehensive radiology report. They repeatedly scroll through CT volume slices multiple times, each time focusing on a specific task and its relevant organs/tissues within a particular CT display window.
Radiologists also access the electronic medical record to help refine their insights from the imaging exam.
This workflow demands that radiologists receive a task command at a time and pay attention to all relevant data to make accurate diagnoses.
To mimic this procedure, we formulate our M3FM in a multimodal question-answering (MQA) framework.
As shown in Fig.~\ref{fig:example}, M3FM encodes multimodal data, incorporates prior knowledge, and predicts the answer to each question.
With prior knowledge, M3FM takes as input the anatomically relevant sub-volume in a specific display window for the given task.
Since different diseases may be distributed in significantly different sizes in CT and correlated with different clinical data portions, M3FM should accommodate various scales of volumetric images and different combinations of multimodal data.
For this purpose, we design a CT Vision Transformer (CTViT) compatible with multiple image sizes as an image encoder of M3FM.
CTViT disentangles the physical size from the image and embeds it as a separate input, which makes M3FM flexibly process different scales of images without losing their physical size information, which is an important factor for specific tasks, see Section~\ref{sec_multimodal}.
We describe all clinical data with free text so that any combinations of multimodal data can be easily input into the model.
Since large-scale self-supervised pre-training has become an essential technique to ensure the performance of large foundation models \cite{liu2019roberta,brown2020language,he2022masked,feichtenhofer2022masked},
we adapted the masked image modeling method \cite{he2022masked,feichtenhofer2022masked} to pre-train our CTViT image encoder with a large number of volumetric CT scans and used the text Transformer pre-trained via masked language modeling \cite{liu2019roberta} as our text encoder.
In our extensive evaluation, we find that by synergizing multimodal data and multitasks, our M3FM significantly improves the performance of the counterpart model trained on single-modality data for individual tasks, and the informative data types for each task can be identified through ablation inference or training. As a foundational model, M3FM can be adapted to boost the performance of new tasks with a small out-of-distribution dataset.

\section{Results}\label{sec2}

\subsection{Multimodal Multitask Datasets}
\label{sec_data}
Figure~\ref{fig:data_all}~(a) shows the general data curation pipeline, including medical tasks of interest definition, task-specific multimodal data collection, multimodal data processing and alignment, and MQA dataset construction.
We target 17 (sub-)tasks related to the LCS process, including 5 tasks for lung nodule detection and characterization, 1 task for cardiovascular disease (CVD) diagnosis,  1 task for CVD mortality prediction, 1 task for lung cancer risk prediction over multiple years, 7 tasks for other chest abnormality exams, 1 task for COVID-19 detection, as well as 1 task for American College of Radiology (ACR) guidelines for Lung CT Screening Reporting and Data System (Lung-RADS) categorization.  COVID-19 detection from CT is included since it remains a global threat \cite{chen2023machine}.
The ground-truth labels come from different information sources, including radiology reports, disease history, pathology test results, follow-up data, death reports, and laboratory test results for different tasks as described in Table~\ref{table_mqa}.
To curate the multimodal datasets, multiple data sources are aligned, including volumetric CT scans, demographics, smoking history, disease history, cancer history, family cancer history, and other task-specific clinical data. In total, 49 different clinical data types are involved in the multimodal datasets for LCS, as described in Table~\ref{table_mqa}.
For each task, one training, one validation, and one or more testing datasets were constructed. An overview of our multimodal multitask dataset is given in Figure \ref{fig:data_all}~(b). The data were collected from different data centers and institutes, including NLST, MIDRC, Wake Forest University School of Medicine (WFUSM), and Massachusetts General Hospital (MGH).
In total, we curated 17 training, 17 validation, and 29 testing datasets for the 17 tasks, with the detailed composition of each dataset in Figure~\ref{fig:data_all}~(c), (d), and (e).
We also collected an out-of-distribution multimodal dataset from WFUSM for transfer learning.
To inspect the modeling ability for textural clinical data, we simulated a dataset for clinical information retrieval, as illustrated in Figure~\ref{fig:example}. 
Since we unify multitask learning with multimodal data in an MQA framework, each dataset consists of task-specific multimodal inputs, questions, and answers. The details for all tasks are summarised in Table~\ref{table_mqa}.

\begin{table*}[ht]
\caption{\textbf{Overview of the Multimodal Question-Answering Datasets.} Note that the nodule sizes are grouped according to the ACR guidelines for Lung-RADS.}
\label{table_mqa}
\includegraphics[width=1\textwidth]{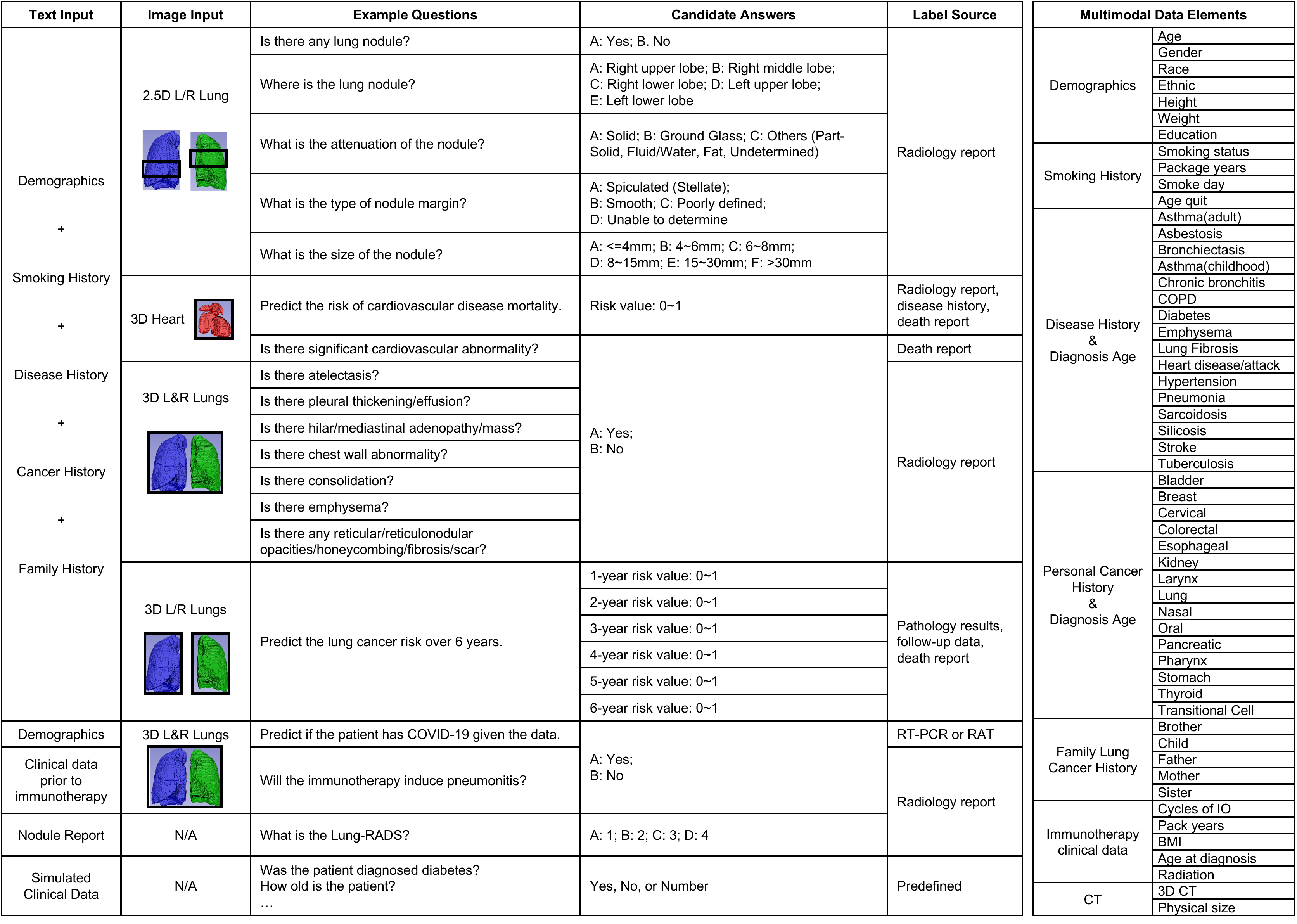}
\end{table*}

\begin{figure}
    \centering
    \includegraphics[width=1.0\textwidth]{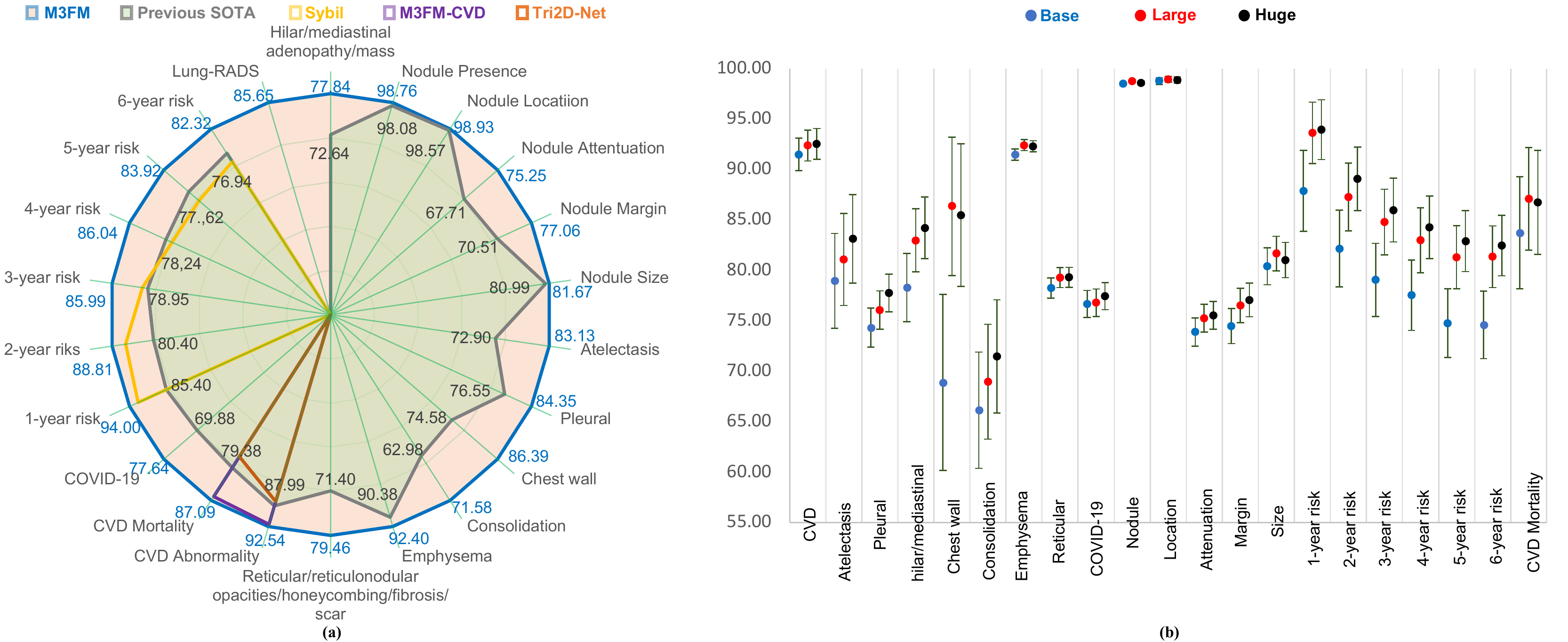}
    \caption{\textbf{Overall performance of M3FM.} (\textbf{a}) Comparison results of the best M3FMs with previous SoTA models in terms of AUC (\%), where the AUC numbers of M3FM and previous SOTA models trained under the same settings are shown and the results of other models can be found in the main text; (\textbf{b}) AUC (\%) results with 95\% CI of three scales of M3FM models. The results demonstrate that M3FM consistently surpasses previous SOTA models across all tasks. Generally, we observed that scaling up the size of the M3FM enhances its performance.}
    \label{fig:ridar}
\end{figure}

As the first data source, we were granted access to all recorded data in NLST.
NLST is a randomized trial for evaluating LCS with 3D LDCT versus 2D chest radiography, demonstrating that screening with LDCT lowered lung cancer mortality by 20\%.
The NLST data were collected from 33 medical institutions, which were randomly indexed without revealing their identifications publicly.
The 26,722 participants in the LDCT screening arm were enrolled from August 2002 through April 2004. 
The participants underwent three screenings at 1-year intervals from August 2002 through September 2007.
The follow-up data were collected until December 31, 2009.
During the whole process, diverse data were recorded, including demographics, smoking history, disease history, multiple CT series with different reconstruction algorithms and associated imaging parameters, key abnormalities in structured reports, pathology test results for lung cancer, follow-up data, and vital status.
Being consistent with clinical practice, we constructed 15 multimodal datasets for 15 tasks, including 5 datasets for predicting the presence of lung nodules and estimating the location, size, margin, and attenuation properties of lung nodules; 7 datasets for identifying chest abnormalities, including atelectasis, pleural thickening/effusion, non-calcified hilar/mediastinal adenopathy/mass, chest wall abnormality (bone destruction, metastasis, etc.), consolidation, emphysema, reticular/reticulonodular opacities/honeycombing/fibrosis/scar; 1 dataset for cardiovascular abnormality diagnosis; 1 dataset for cardiovascular mortality risk prediction; and 1 dataset for lung cancer risk prediction within from 1 to 6 years. Each dataset was randomly split into training, validation, and test datasets. The patient-wise information in the validation and test datasets is not leaked to the training datasets across all tasks.
From NLST, we included 125,090 effective volumetric chest CT scans of 26,254 patients.

The second data source is the Medical Imaging and Data Resource Center (MIDRC), a collaboration of leading medical imaging organizations launched in August 2020 as part of NIBIB's response to the COVID-19 pandemic.
We were granted to access all CT series with the associated clinical data. The ground-truth labels for COVID-19 are determined by either the Reverse Transcription Polymerase Chain Reaction (RT-PCR) or the Rapid Antigen Test (RAT).
From MIDRC, we retrieved 35,730 volumetric chest CT series of 7,609 patients being scanned from 2011 to 2021.
The patient data were randomly split into the training, validation, and test datasets.

All CT scans from NLST and MIDRC excluding those in any test datasets were combined as a CT pretraining dataset, comprised of 151,569 CT scans in total.
To inspect if the clinical data are effectively encoded, we constructed a clinical question-answering dataset to retrieve key information from the textual clinical data.
The integration of all the above-curated datasets is called OpenM3Chest, which will be released upon this paper is accepted.
To our best knowledge, OpenM3Chest will be the first-of-its-kind publicly available MQA dataset and also the largest 3D CT dataset for training AI models.

To test the generalizability and adaptability of M3FM, we independently collected two multimodal multitask datasets from the third and fourth data sources, i.e., WFUSM and MGH, respectively.
Each dataset includes CT scans, radiology reports, demographics, smoking history, disease history, personal cancer history, family lung cancer history, and pathology test results for lung cancer.
The MGH and WFUSM review boards approved the analysis of all these multimodal data and tasks.
Based on the radiology reports and the pathology test results, we constructed 7 datasets from WFUSM and 6 datasets from MGH for independent evaluation as shown in Figure~\ref{fig:data_all}.
Specifically, we collected 8,053 patient data at WFUSM from 2015 to 2023, all with radiology reports and 1,800 of them with LDCT and multimodal information.
We collected 904 patient data with multimodal data shown in Table~\ref{table_mqa} at MGH from 2016 to 2021.
The Lung-RADS dataset from WFUSM was randomly split into training, validation, and test datasets, which were defined to classify the text descriptions into the Lung-RADS category. All other datasets of WFUSM and MGH were used for testing.
To evaluate the adaptability of our M3FM, we collected an out-of-distribution multimodal dataset for non-small cell lung cancer (NSCLC) immunotherapy prognosis from WFUSM. This dataset consists of 90 patient data, including the target label indicating if the patient was diagnosed with immune checkpoint-inhibitor-induced pneumonitis after immunotherapy, the CT scans before immunotherapy, and the clinical variables including the total cycles of Immuno-Oncology (IO), smoking information of pack years, Body Mass Index (BMI) at diagnosis, age, and if the patient received radiation prior to immunotherapy. Among the 90 patients, 49 patients developed immune checkpoint-inhibitor-induced pneumonitis, and the other patients were used as the control group.

The details on the multimodal data processing and alignment and the MQA dataset construction are described in Section~\ref{sec_method_data}.

% \subsubsection{ChestFM Architecture}

\subsection{M3FM Setting New Records}
Figure~\ref{fig:ridar}~(a) summarizes the key results of M3FM against the SOTA models~\cite{mikhael2023sybil, chao2021deep} on the OpenM3Chest dataset.
Overall, M3FM sets new records for all tasks, significantly outperforming the SOTA models for most of these tasks.
We used the Area Under the receiver operating characteristic Curve (AUC) and the 95\%  two-sided Confidence Intervals (CI) of AUC values proposed by Hanley and McNeil~\cite{hanley1982meaning} as the evaluation metrics.
For a fair comparison, we retrained the Sybil model~\cite{mikhael2023sybil}, denoted as Sybil$^*$, for lung cancer risk prediction without using the costly bounding box annotations but predicting lung cancer risks by merging the separate results of left and right lungs. 
It is observed that Sybil$^*$ achieved inferior results for 1$\sim$2-year risk prediction but superior results for 3$\sim$6-year risk prediction in comparison with the results obtained using the original Sybil model.
Without using any bounding box, our M3FM achieves an AUC of 94.00\% (95\% CI, 91.19-96.98), 88.81\% (95\% CI, 85.67\%-91.95\%), 85.99\% (95\% CI, 82.88\%-89.1\%), 86.04\% (95\% CI, 83.10\%-88.98\%), 83.92\% (95\% CI, 80.98\%-86.85\%), 82.32\% (95\% CI, 79.36\%-85.29\%) for lung cancer risk prediction over six years, outperforming both Sybil$^*$ and original Sybil models by the margins of 5\% to 9\% and 2\% to 11\%, respectively.
For CVD diagnosis and CVD mortality prediction, we compared the results on both the original dataset~\cite{chao2021deep} and our OpenM3Chest dataset.
M3FM achieves an AUC of 92.84\% (95\% CI, 91.36\%-94.33\%) for CVD diagnosis and an AUC of 89.04\% (95\% CI, 84.27\%-93.81\%) for CVD mortality prediction on the OpenChest dataset, outperforming the previous models (Tri2D-Net~\cite{chao2021deep}) by 5\% and 9\% respectively, and achieves an AUC of 92.03\% for CVD diagnosis and 86.14\% for CVD mortality prediction on the datasets constructed in~\cite{chao2021deep}, outperforming the previous models (Tri2D-Net) by 5\% and 10\% respectively.
For several tasks including nodule detection, nodule localization, nodule size prediction, and emphysema detection, M3FM improves the results by various degrees up to 3\% of AUC.
For all the other tasks, M3FM significantly improves the performance from $\sim$5\% to $\sim$10\%.
To study the scalability of M3FM, we trained three versions of M3FM, consisting of 257M (M3FM-Base), 502M (M3FM-Large), and 865M (M3FM-Huge) trainable parameters respectively. The results of these three models are summarized in Figure~\ref{fig:ridar}~(b).
Overall, with a larger model size, the performance becomes better, especially from M3FM-Base to M3FM-Large. This trend is consistent with the well-known scaling law \cite{kaplan2020scaling} in the field of foundation models.

\begin{table*}[h]
\caption{\textbf{Comparison of M3FM Variants on the OpenM3Chest Dataset.} SM denotes single-modality, MM signifies multi-modality, ST represents single-task, and MT indicates multitask. Gray-colored values detail specific task categories. Analysis of AUC (\%) results with a 95\% CI revealed that multi-modality models outperform single-modality models in 14 of 21 tasks, as highlighted by the \underline{underlined} values in the comparison between M3FM-SM-ST and M3FM-MM-ST. Furthermore, multitask models surpass single-task models in 17 of 22 tasks, as emphasized by the \textbf{bold} values in the comparison between M3FM-MM-ST and M3FM-MM-MT.}
\label{table_mt}
\includegraphics[width=1\textwidth]{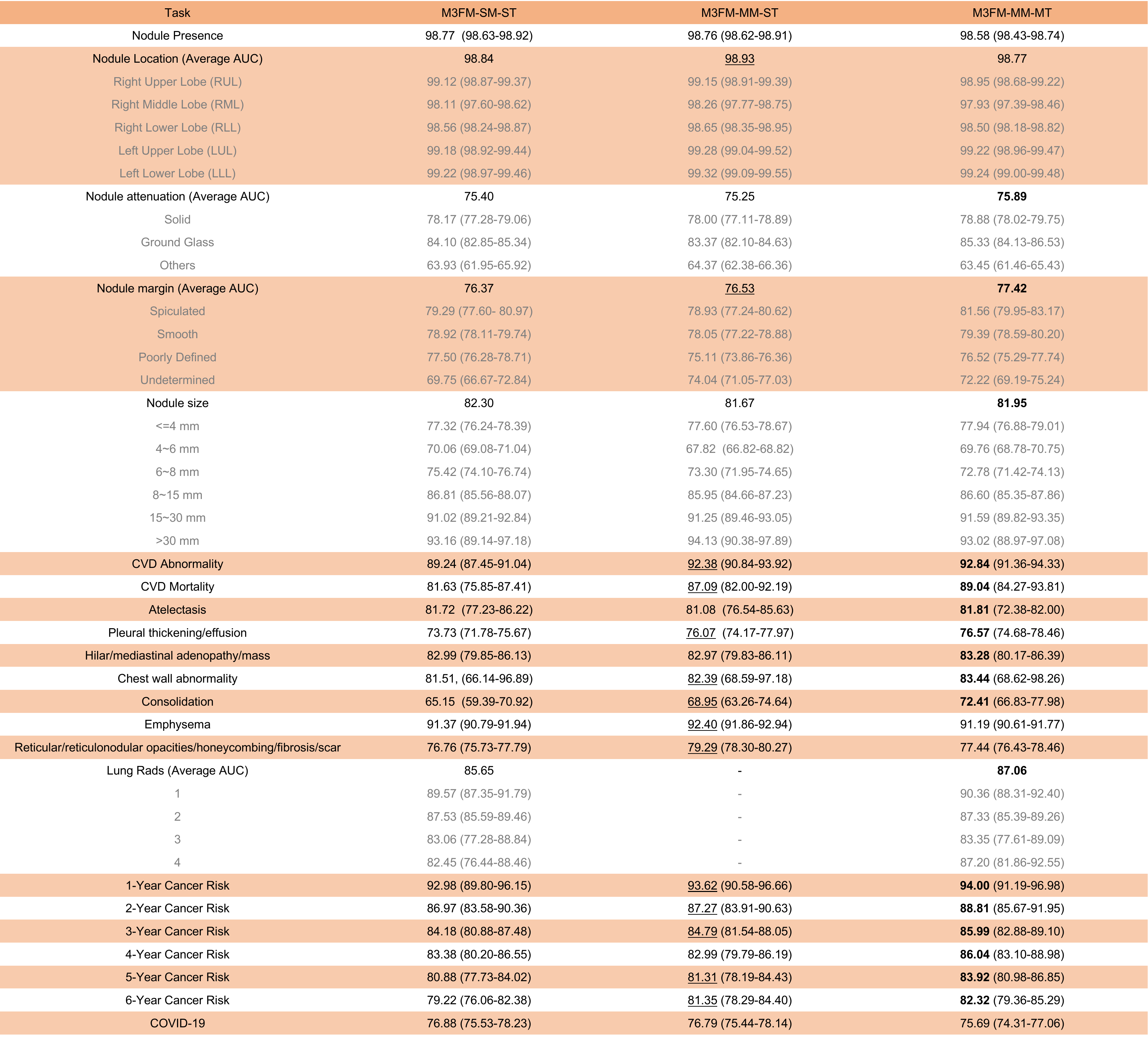}
\end{table*}

\subsection{M3FM Encoding Multimodal Data and Synergizing Multiple Clinical Tasks}
\label{sec_multimodal}

Table~\ref{table_mt} compares the results of the single-modality single-task, multi-modality single-task, and multi-modality multitask M3FM-Large models.
First, the single-modality single-task models were trained and evaluated on LDCT data only and denoted by M3FM-SM-ST, while the multi-modality and single-task models were trained and evaluated on multimodal data and denoted by M3FM-MM-ST.
Overall, the multimodal information improves the prediction results for multiple tasks.
In particular, M3FM-SM-ST achieves an AUC of 81.63\% (95\% CI, 75.85\%-87.41\%) for CVD mortality prediction while the M3FM-MM-ST model achieves an AUC of 87.09\% (95\% CI, 82.00\%-92.19\%), which represents a 5.46\% improvement.
While M3FM-SM-ST achieves an AUC of 89.24\% (95\% CI, 87.45\%-91.04\%) for CVD diagnosis, the M3FM-MM-ST model achieves an AUC of 92.38\% (95\% CI, 90.84\%-93.92\%), i.e., a 3.14\% improvement.
Similarly, M3FM-SM-ST achieves an AUC of 65.15\%  (95\% CI, 59.39\%-70.92\%) for consolidation detection, and the M3FM-MM-ST model achieves an AUC of 68.95\% (95\% CI, 63.26\%-74.64\%), a 3.80\% improvement.
Also, M3FM-SM-ST achieves an AUC of 76.76\% (95\% CI, 75.73\%-77.79\%) for reticular/reticulonodular opacities/honeycombing/fibrosis/scar detection, and the  M3FM-MM-ST model achieves an AUC of 79.29\% (95\% CI, 78.30\%-80.27\%), a 2.53\% improvement.
It is further observed that M3FM-MM-ST models produce slightly improved or comparable results in comparison with M3FM-SM-ST for the other tasks.
Then, we compared the multimodal multitask model (M3FM-MM-MT) and multimodal single-task models (M3FM-MM-ST).
Impressively, training on multiple tasks, M3FM-MT-MM outperforms the M3FM-ST-MM for 17 out of 22 (sub)-tasks.

\begin{table*}[ht]
\caption{\textbf{Evaluation of clinical elements in CVD tasks.} In terms of AUC (\%) results with 95\% CI,  we observed that multi-modality models are significantly better than single-modality models for CVD tasks. In particular, the results revealed that the history of heart disease/attack, hypertension, diabetes, and stroke are highly informative in CVD diagnosis and CVD mortality risk prediction.}
\label{table_acvd}
\includegraphics[width=1\textwidth]{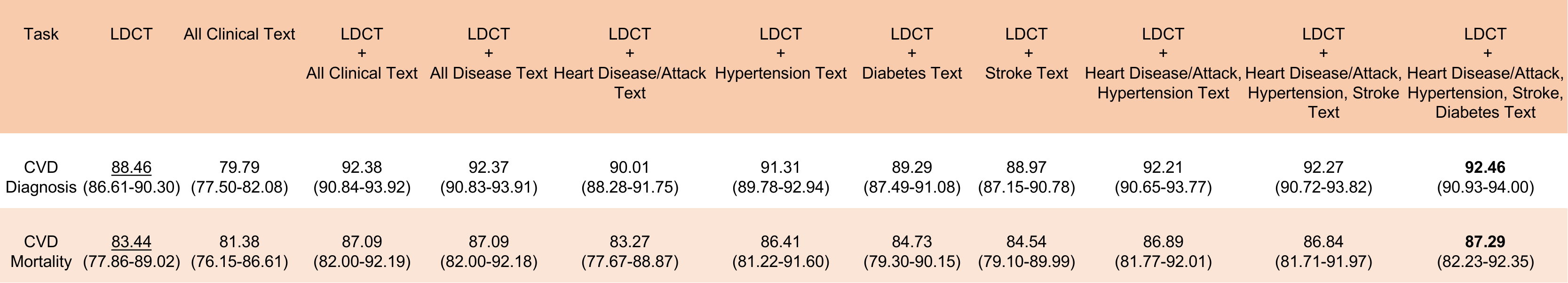}
\end{table*}

\begin{table*}[ht]
\caption{\textbf{Evaluation of clinical elements in lung cancer risk prediction.} In terms of AUC (\%) and 95\% CI results, we observed that the combination of LDCT and demographic achieved the best results. Nevertheless, clinical text data did not significantly enhance the lung cancer risk prediction.}
\label{table_acancer}
\includegraphics[width=1\textwidth]{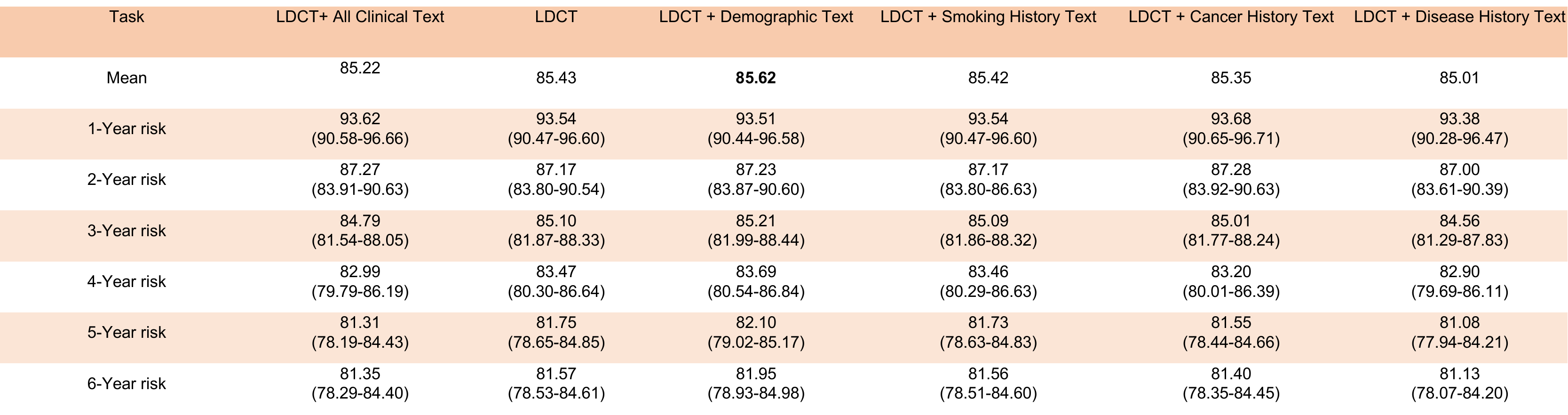}
\end{table*}

\begin{figure}
    \centering
    \includegraphics[width=1\textwidth]{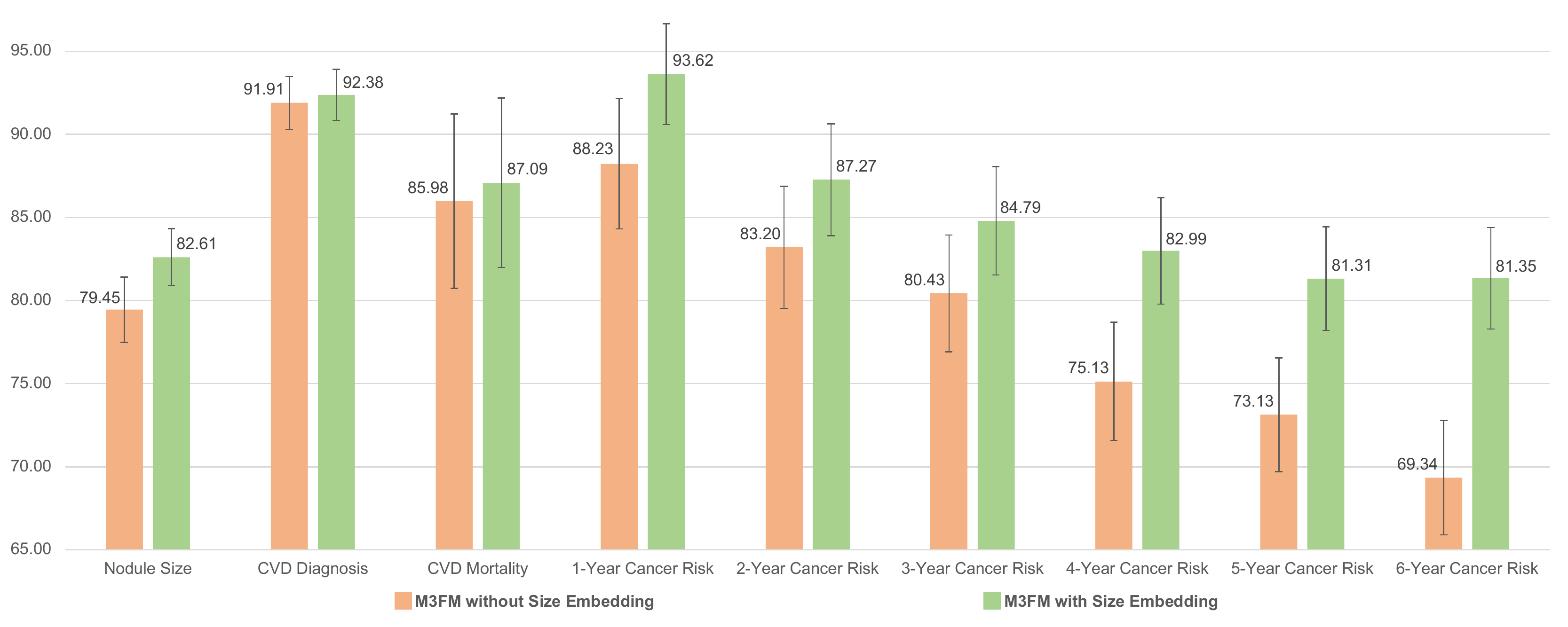}
    \caption{\textbf{Evaluation of physical size embedding in CT imaging.} This figure presents the AUC~(\%) and 95\% CI results for M3FM models with and without the embedding of CT voxel sizes across various tasks. The inclusion of physical size information significantly enhances performance in lung cancer risk prediction, cardiovascular disease (CVD) diagnosis, CVD mortality risk estimation, and nodule size characterization.}
    \label{fig:size}
\end{figure}

% \begin{table*}[ht]
% \caption{Effects of embedding physical CT voxel size for different tasks in terms of AUC (\%) and 95\% CI.}
% \label{table_size}
% \includegraphics[width=1\textwidth]{size.pdf}
% \end{table*}

\begin{figure}
    \centering
    \includegraphics[width=1\textwidth]{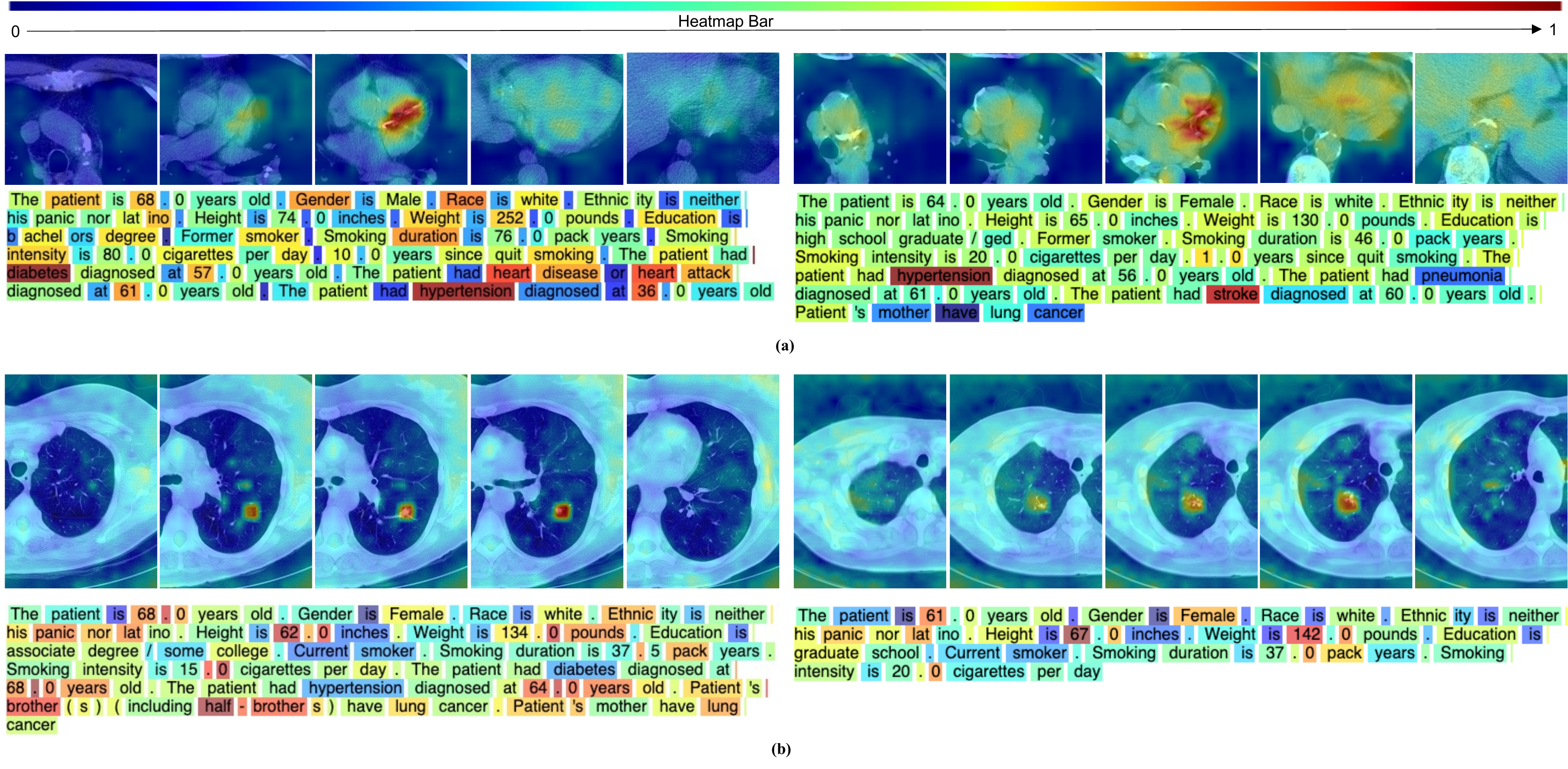}
    \caption{\textbf{Qualitative inspection of the task encoder.} (\textbf{a}) Attention map visualization of the task encoder for two CVD diagnosis examples; (\textbf{b}) Attention map visualization of the task encoder for two lung cancer risk prediction examples. The exemplar results show that the task encoder has a certain ability to reveal the relevance between the model outcomes and multimodal elements. Specifically, CVD diagnosis correlates with calcification regions and the patient's history of heart disease, hypertension, diabetes, and stroke. Lung cancer risk is associated with the lung nodule region as well as demographic factors and familial lung cancer history. In row (\textbf{a}), the two cases were reported with significant CVD abnormalities. In row (\textbf{b}), the pathology test results confirmed the two patients being diagnosed with lung cancer within one year following their LDCT lung cancer screenings.}
    \label{fig:att}
\end{figure}

\subsection{M3FM Identifying Clinically Informational Elements}

Since M3FM accommodates any combination of multimodal datasets in the training and inference stages, we investigated the application of M3FM to analyze the synergy between clinical elements and tasks by observing the effects of different input combinations on the model outcomes.
Table~\ref{table_acvd} presents the ablation results using different combinations of multimodal data for CVD diagnosis and mortality prediction.
M3FM using all multimodal inputs improved the AUC by 3\% $\sim$ 4\% relative to the results using LDCT only and by 12\% and 5\% over that using clinical data only for CVD diagnosis and mortality prediction respectively.
Furthermore, the M3FM results show that the disease histories of heart disease or heart attack, hypertension, stroke, and diabetes consistently boosted the AUC results by gradually adding them into different input combinations for CVD diagnosis and mortality prediction.
Table~\ref{table_acancer} shows the lung cancer risk prediction results using different inputs, where we found that demographic information slightly improved the AUC results.

Then, we evaluated if M3FMs could effectively encode the physical size information.
The ablation results in Figure~\ref{fig:size} show that the embedded physical size information of LDCT improves the AUC results for multiple tasks. The physical size information boosts the AUC of 1$\sim$6-year lung cancer risk prediction by 5\%, 4\%, 4\%, 7\%, 8\% and 12\%, respectively.
The physical size information also improves AUC results of the nodule size characterization, CVD diagnosis, and CVD mortality prediction, by 0.71\%, 0.47\%, and 1.11\% respectively.

We quantitatively evaluated the correction of different clinical elements with model outputs by visualizing the attention maps of the last task attention block in M3FM.
Figure~\ref{fig:att} visualizes the attention heat maps on the selected CT slices and the text tokens of individual patients with CVD or lung cancer risks.
In CVD diagnosis, the coronary artery calcification areas were highlighted in the LDCT attention heat maps, and the patients' disease histories of diabetes, heart disease or heart attack, hypertension, and stroke are highly relevant among text tokens, which is consistent with the quantitative results in Table~\ref{table_acvd}.
In predicting lung cancer risks, the lung nodules in LDCT images were localized in the heat maps, and the text tokens related to demographic and family lung cancer histories are more correlated to the model outputs.

% \begin{table*}[ht]
% \caption{Evaluation results on the MGH and WFUSM datasets in terms of AUC and 95\% CI.}
% \label{table_independent}
% \includegraphics[width=1\textwidth]{independent.pdf}
% \end{table*}

\begin{figure}
    \centering
    \includegraphics[width=1\textwidth]{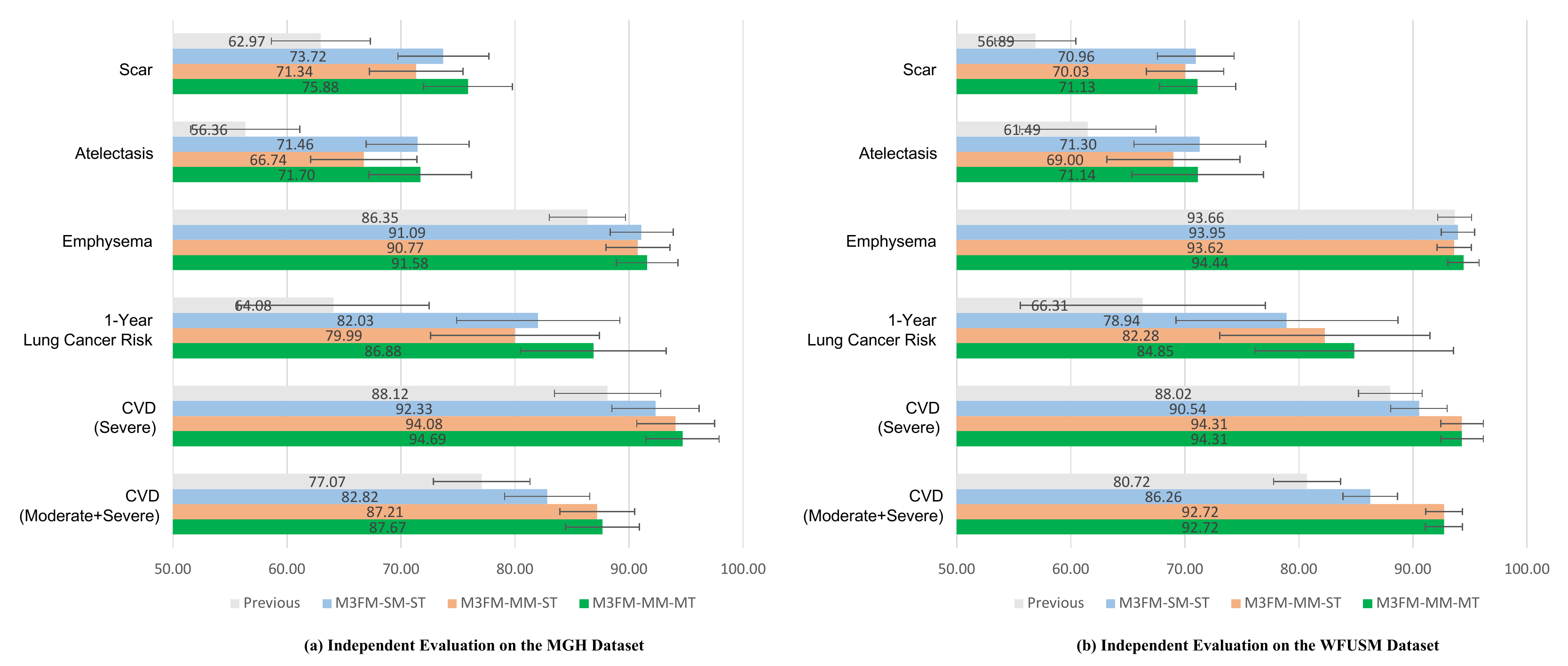}
    \caption{\textbf{Evaluation of different models on independent datasets.} Evaluation results of the previous model and M3FM variants on the (\textbf{a}) MGH and (\textbf{b}) WFUSM datasets in terms of AUC (\%) and 95\% CI. The results show that M3FM has significantly better generalizability than the previous models, especially boosting the 1-year lung cancer risk prediction by up to 20\%. Furthermore, these independent evaluation results confirmed the efficacy of multimodal and multitask modeling.}
    \label{fig:independent}
\end{figure}

\subsection{M3FM Improving Generalizability}

We evaluated the generalizability of M3FMs on the independently collected datasets, with the comparison results on the MGH and WFUSM datasets shown in Figure~\ref{fig:independent}.
For the CVD diagnosis task, we constructed two datasets, which regard moderate and severe CVD as positive and severe CVD only as positive, respectively.
On the two MGH CVD datasets, the multimodal multitask model (M3FM-MM-MT) improved the AUC by 10.60\% and 6.57\% relative to the previous model, improved the AUC by 4.85\% and 2.36\% relative to the single-modality single-task model (M3FM-SM-ST), and also achieved slight AUC improvements relative to the multi-modality single-task model (M3FM-MM-ST). Relative to M3FM-SM-ST, the M3FM-MM-ST model improved the AUC by 4.39\% and 1.75\% on the two CVD MGH datasets.
For the 1-year lung cancer risk prediction on the MGH dataset, the M3FM-MM-MT model improved the AUC by 20.80\% than the previous model under the same experimental settings without using any bounding box annotations, improved the AUC by 4.85\% than M3FM-SM-ST, and improved the AUC by 6.89\% than M3FM-MM-ST.
On the MGH emphysema, atelectasis, and reticular opacities/honeycombing/fibrosis/scar datasets, the M3FM improved the AUC by 5.23\%, 14.34\%, and 12.91\% than the previous model, and also achieved AUC improvements by 0.24\% $\sim$ 4.96\% than M3FM-SM-ST and M3FM-MM-ST.
For the CVD tasks on MGH datasets, M3FM-MM-MT improved the AUC by 12\% and 6.29\% than the previous model, improved the AUC by 6.46\% and 3.77\% relative to M3FM-SM-ST, and had the same results with M3FM-MM-ST.
For the 1-year lung cancer risk prediction on the MGH dataset, the M3FM-MM-MT model improved the AUC by 18.54\% relative to the previous model under the same experimental settings without using any bounding box annotations, improved the AUC by 5.91\% than M3FM-SM-ST, and improved the AUC by 2.57\% than M3FM-MM-ST; and M3FM-MM-ST improved the AUC by 3.24\% than M3FM-SM-ST.
On the WFUSM emphysema, atelectasis, and reticular opacities/honeycombing/fibrosis/scar datasets, the M3FM-MM-MT model improved the AUC by 0.78\%, 9.65\%, and 14.24\% than the previous model.

% More evaluation results on WFUSM datasets will be reported here.

\begin{figure}
    \centering
    \includegraphics[width=1\textwidth]{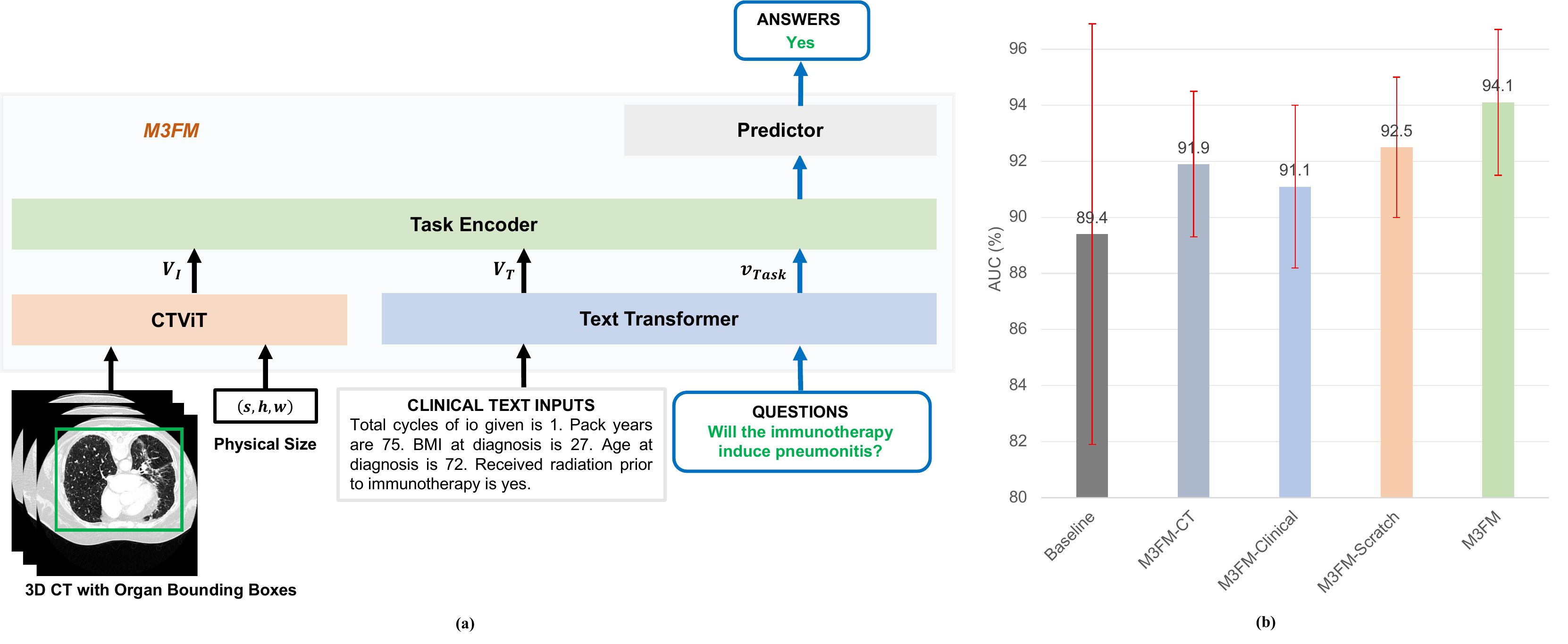}
    \caption{ \textbf{Transfer learning with M3FM.} (\textbf{a}) The same M3FM architecture was fine-tuned to perform the new immunotherapy prognosis task with different CT and clinical inputs; (\textbf{b}) Results of immunotherapy-induced pneumonitis using different methods. These results demonstrate that M3FM is adaptable to enhance the out-of-distribution task.}
    \label{fig:immu}
\end{figure}

\subsection{M3FM Enhancing Out-of-Distribution Multimodal Analysis}

As a foundational model, we evaluated if M3FM facilitates out-of-distribution multimodal modeling. Here we fine-tuned the M3FM to predict immunotherapy-induced pneumonitis from volumetric CT prior and the selected clinical data related to immunotherapy as described in Sub-section~\ref{sec_data}. We used the method developed in WFUSM as the baseline and compared different fine-tuned variants of the M3FM in terms of the average AUC and its standard deviation of five-fold cross-validation. The baseline model used in WFUSM had 89.4\% $\pm$ 7.5\% AUC by merging all radiomics and clinical features. 
Specifically, the baseline model was based on a nomogram to predict immunotherapy outcomes using features extracted from radiomic algorithms, a pre-trained ViT-base model, and clinical records. After feature selection, 20 radiomic features, 20 deep features, and 17 clinical features were used for the nomogram.
The best result of our fine-tuned M3FMs was 94.1\% $\pm$ 2.6\% of AUC, which achieved a 4.7\% improvement.
The M3FM-CT model using CT data only had an AUC of 91.9\% $\pm$ 2.6\%. The M3FM-Clinical model using clinical text only had a 91.1\% $\pm$ 2.9\% AUC. The M3FM-Scratch without pretraining achieved 92.5\% $\pm$ 2.5\% of AUC.

\section{Discussions}

The contributions of the proposed M3FM can be summarized in two main aspects.
First, as the first-of-its-kind SOMAI model for LCS, M3FM effectively encodes multimodal medical data including arbitrary combinations of multi-scale 3D tomographic images and various clinical data, and flexibly performs multiple tasks via free-text prompting.
In particular, our CT Vision Transformer (CTViT) is a unique component designed to perceive 3D CT images.
CTViT can flexibly process multiple image sizes through our multi-scale linear tokenizer and disentangled physical size embedding designs.
We developed the corresponding self-supervised learning training algorithm that facilitates the pre-training of the multi-scale CTViT on large 3D CT datasets.
To make M3FM scalable across multiple tasks, we designed a distributed task-parallel training strategy, which assigns a single task to each device while allowing different devices to process different inputs/outputs for multitask parallel optimization.
Second, we presented the whole workflow for the SOGMAI model development, from clinical multitask definition to multimodal data curation, from radiologists’ reading procedure to the unified MQA framework, and from self-supervised pre-training to synergistic multitasking with high-dimensional multimodal data.
In particular, our MQA framework is akin to how medical professionals perform multiple tasks while considering multimodal data, naturally allowing unified training and interactive inference.
Importantly, the whole pipeline is designed with scalability, allowing M3FM to be readily scaled up by integrating more training datasets and undertaking a broader range of clinical tasks.

The M3FMs set the new performance records on all tasks of interest on our curated large-scale OpenM3Chest datasets collected from multiple medical institutes.
The M3FMs significantly outperformed the previous models, which mainly focus on task-specific and single modality.
Promisingly, our experimental results affirm that the larger M3FM produces better outcomes in medical multimodal multitask settings.
These positive outcomes underscore the importance of systematically collecting and curating large-scale, multimodal, multitask datasets to enhance the precision of medical AI applications.
All these findings on clinical datasets highlight the potential of SOGMAI.

The M3FMs effectively encode multimodal data and flexibly synergize different medical tasks.
Importantly, different tasks may have different combinations of multimodal inputs and varying output formats.
Generally, our findings indicate that multimodal modeling can enhance performance on specific tasks, while multitask learning tends to improve outcomes across a broader range of tasks.
Still, some tasks may only depend on single modality data; e.g., lung nodule detection and characterization do not benefit from clinical datasets more than LDCT scans in our experiments.
The overall results from multimodal and multitask modeling efforts are encouraging, indicating that M3FM possesses the potential to encompass a wider array of medical tasks and datasets for unified and improved medical AI applications.

Nevertheless, there are certain limitations to current M3FM results. The above evaluation was retrospective and offline rather than in a prospective, real-world reporting and patient management environment.
In current clinical systems, real-time multimodal data integration could pose logistical challenges, particularly when managing information streams across multiple, distinct yet interconnected interfaces such as electronic medical records, radiology information systems, and picture archiving and communication systems. We did not test the clinical impact of M3FM either in radiology or post-radiology care scenarios. Likewise, we did not evaluate the most effective method of information display to improve decision-making with multimodal information without inundating physicians and compromising their workflow efficiency. 
While our experimental results affirm that larger-scale models yield better outcomes in medical multimodal multitask performance, the performance gain when upgrading from M3FM-Large to M3FM-Huge is less impressive than when upgrading from M3FM-Base to M3FM-Large. We believe that this limited improvement could be substantially attributed to the size and quality of the current datasets, and with even larger and better datasets we expect to have superior performance, by the scaling laws.
Within the scope of these limitations, there are opportunities for clinical impacts of M3FM. With regulatory clearance and integration with real-time clinical workflows, M3FM can provide a dynamic, and customizable dashboard for information summary and decision-making. For example, during the reporting of lung findings, it would be helpful to have a display of M3FM-derived results, previously reported lung findings, and lung-specific clinical findings (such as Chronic Obstructive Pulmonary Disease (COPD), prior lung nodules, and LungRADS categories), and during cardiovascular “field” reporting, to have a display of M3FM-based estimates, cardiac risks factors from past medical history and laboratory/ECG/echocardiography reports. Beyond LCS, such a dashboard can help in various other ways; for example, in cancer staging, we can access not only M3FM outputs but also medical, radiation, and surgical treatments when determining the best management.

As discussed in \cite{cases}, there are specific concerns for AI in medicine, such as generalizability, explainability, adaptability, etc. This study has demonstrated initial efforts in addressing AI-specific concerns.

The M3FMs show better generalizability to independent WFUSM and MGH datasets than task-specific models.
It is worth mentioning that the independent evaluation is prospective in terms of the date of data collection.
Our experimental results show that the multimodal multitask M3FM models achieved consistently and significantly better results by up to 20+\% than the previous models trained in the same settings.
However, for some tasks, the multi-modality modeling may decrease the generalizability relative to the single-modality modeling.
This could be due to the variability in data collection procedures and standards.
Thus, it is important to design a standardized and robust data resourcing and collecting pipeline.
In all the experiments, multitask learning can consistently improve generalizability.

The M3FMs are capable of identifying informative clinical elements both quantitatively and qualitatively, which offers certain explainability. It is achieved with the MQA framework and the attention mechanism. Specifically, 
the MQA framework naturally allows users to examine the response changes to different combinations of imaging and clinical data, and thus, the informative clinical elements can be identified as those contributing to statistically high prediction accuracy.
Our M3FMs have uncovered a strong positive correlation between CVD diagnosis, CVD mortality prediction, and the historical presence of heart disease/attacks, hypertension, stroke, and diabetes through ablation inference.
Assuming this discovery is not a piece of common knowledge, M3FM would lead to new biomedical findings.
Quantitatively, attention maps can be visualized for both image and text inputs through the attention mechanism, illuminating the elements that correlate with predictions. This visualization offers certain the interpretability of M3FM models in understanding the relationships between clinical data and diseases.

The M3FM has the adaptability to significantly improve multimodal modeling for out-of-distribution tasks through transfer learning. A key feature of foundation models is their ability to aid tasks beyond those defined by the training datasets. In this study, we finetuned our M3FMs for immunotherapy prognosis prediction, an out-of-distribution task characterized by entirely different clinical inputs. Our experiments demonstrated that the pre-trained M3FM model substantially enhances the new task performance on a relatively small dataset. This capability is particularly valuable when expanding some clinical datasets is challenging due to data rarity and associated costs.

In conclusion, the unified architecture and exceptional performance of the M3FMs herald a promising avenue for leveraging multimodal data and clinical multitasks in developing AI-empowered, specialty-oriented superior healthcare solutions. Within the scope of LCS in particular, we see the potential to translate the M3FM model on our collaborative clinical sites, broaden and refine LCS implementation, and ultimately reduce lung cancer mortality. Hopefully, our M3FM system would become an effective platform to accommodate more medical tasks with diverse multimodal data combinations, from specialized to increasingly more generalized medical AI models.

% \begin{figure}
%     \centering
%     \includegraphics[width=1\textwidth]{dataflow.pdf}
%     \caption{General Workflow for Medical Multimodal Multitask Dataset Construction.}
%     \label{fig:df}
% \end{figure}

\section{Methods}\label{sec11}
\label{sec_method}

\subsection{Medical Multimodal-Multitask Dataset Construction}
\label{sec_method_data}
Figure~\ref{fig:data_all}~(a) presents our general workflow for constructing the multimodal multitask medical datasets, which consists of four main steps: (1) medical tasks of interest definition; (2) task-specific multimodal data collection; (3) multimodal data processing and alignment; and (4) MQA construction.
The details for the first two steps are described in Section~\ref{sec_data}.
Here we introduce the third and fourth steps.

The multimodal data processing is to select qualified multimodal data and prepare them for the alignment, including the CT data processing, clinical data processing, and ground-truth label calculating.
In CT data processing, we localize the sub-volumes that mainly contain the task-relevant regions in the 3D CT using a segmentation model~\cite{tseg}. Specifically, we segment three parts, i.e., left lung, right lung, and heart regions consisting of the myocardium, left/right atrium, left/right ventricle, and pulmonary artery. It is worth underlining that the precision of the segmentation results does not need to be extremely high. Our primary objective is to utilize rectangular boxes to wrap the segmented areas, ensuring that task-relevant sub-volumes are included and extraneous regions are disregarded.
We excluded the CT series having less than 64 axial slices in all collected datasets.
For each CT series, the reconstruction voxel sizes are prepared in the axial, coronal, and sagittal dimensions, which will be used as input to the CTViT.
The clinical data processing is to represent various combinations of clinical data within free text. We have established a specific sentence format for each clinical element, as detailed in Table S1 of the supplementary. The final free-text clinical data for each examination is constructed by aggregating the sentences corresponding to all available and positive clinical data, as illustrated by the text inputs in Figures~\ref{fig:example} and \ref{fig:immu}.
In ground-truth label calculating, we first extract task-specific label information from different sources (see Table~\ref{table_mqa} for specific label sources of each task) and then combine all information to calculate the label. The details for each task ground-truth label calculation are described in Table S2 of the supplementary.
Next, we align the clinical data in free text, the CT data with segmented parts and physical size, and the calculated labels for all exams in each task.
In particular, each task anatomically corresponds to the segmented CT sub-volume. The key principle here is to remove the irrelevant image regions to reduce the computation cost while keeping the original resolution without losing the information.
The task-specific CT sub-volumes are illustrated in the image input column of Table~\ref{table_mqa}.
Specifically, 2.5D left or right lung sub-volume is used as the image input for lung nodule detection and characterization. We fixed the number of slices to 16 for each input considering that a lung nodule is usually tiny relative to the whole lungs. For the nodule-presented case, the location labels of the left/right lung, slice number, and the bounding box coordinates are used to crop the target sub-volume. For the non-presented nodule case, the input sub-volume is randomly cropped within the segmented lung regions.
The CVD tasks use a 3D rectangle box wrapping all heart regions as the image input.
For lung cancer risk prediction, we separately input the left or right lung in a 3D rectangle box.
In the training datasets, the location of the left and/or right lung where lung cancer is presented is required to align each lung with the risk labels.
In the validation and test datasets, the location of lung cancer is not required as the ground-truth labels are patient-level and the predicted lung cancer risks are the maximum scores between two lungs for each patient.
For all other chest disease diagnosis tasks, a 3D rectangle box wrapping both lungs is used as the image input.
Subsequently, the MQA construction is to create questions and answers for each specific task with the aligned multimodal data, and the resulting MQA datasets define the model's input and output formats. In Table~\ref{table_mqa}, one example question and the corresponding answer candidates are presented for each task. In the training stage, ten different questions for each task were used as shown in Table S3 of the supplementary.
Note that there is no need for extra labeling efforts by radiologists across the whole workflow so that large-scale medical multimodal multitask datasets can be systematically and cost-effectively constructed.

\begin{figure}
    \centering
    \includegraphics[width=1\textwidth]{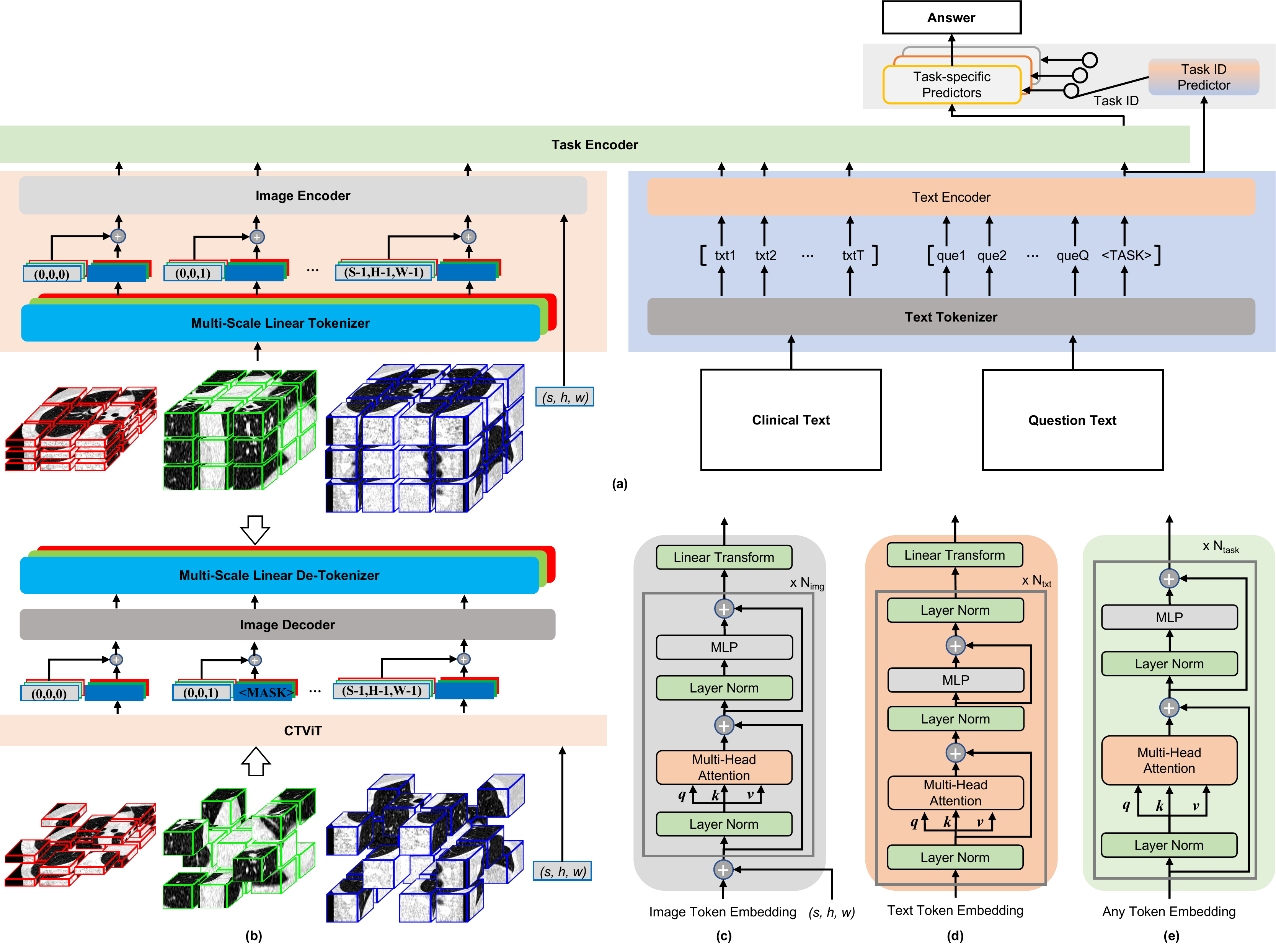}
    \caption{\textbf{M3FM architecture.} (\textbf{a}) the overall M3FM architecture with CTViT, text transformers, associated encoders, and predictors; (\textbf{b}) CTViT pretraining; (\textbf{c}) image encoder; (\textbf{d}) text encoder; and (\textbf{e}) task encoder.}
    \label{fig:method}
\end{figure}

\subsection{M3FM}

\subsubsection{Overall architecture}

Our medical multimodal multitask foundation model is designed to effectively encode multimodal data and flexibly perform multitasks via text prompting in a unified and scalable fashion.
As shown in Figure~\ref{fig:example}, M3FM consists of the four main components: CTViT, text Transformer, task encoder, and predictors. The key details of each component are given in Figure~\ref{fig:method}.
CTViT takes volumetric CT images of varying sizes as inputs, extracts multi-scale image patches from them, and computes discriminative features of these patches.
The text Transformer produces the embeddings of clinical text and the embeddings of textual questions respectively.
Given any combination of image, text, and task token embeddings, the task encoder extracts the task-specific features corresponding to the special $<TASK>$ token.
Finally, the task-specific predictor outputs the final answer from the task-specific features of the integrated multimodal data.
In the following, we will describe each of these components in detail.

\subsubsection{CTViT}
\label{sec_imgencoder}
CTViT extracts embedding features of multi-scale 3D CT volumes with physical size awareness.
CTViT has two parts: a multi-scale CT tokenizer and an image encoder.
To process a 3D CT scan, we divide each image volume into non-overlapped 3D patches as in \cite{niu2022unsupervised}. Each 3D patch is referred to as an image token.
Since various diseases are at different scales in the CT images, 
we design a multi-scale CT tokenizer, which consists of multiple linear embedding layers corresponding to different sizes of image patches, as shown in Figure~\ref{fig:method}~(a).
Each embedding layer has a linear transformation and a set of learnable positional embeddings. Each image token embedding is the sum of its linear transformation and the positional embedding.
All sizes of image tokens are mapped to the same image embedding space.
Inspired by \cite{feichtenhofer2022masked}, we decompose the 3D position embedding into two parts indexing in-plane and through-plane positions respectively. In other words, we have two positional embeddings: one for the 2D space within each slice and the other for the 1D range of slice position. The 3D positional embedding is the sum of them. By doing so, the number of learned parameters can be reduced.
Figure~\ref{fig:method}(c) shows the image encoder in detail.
Different CT scans may have different physical sizes specific to the individual patient size.
The physical size is an important factor in some clinical tasks.
Thus, we encode the physical size for image tokens with sine-cosine functions of different frequencies and add it to the image token embedding.
Then, the image encoder was implemented as the plain ViT~\cite{dosovitskiy2020image} that consists of multiple self-attention Transformer layers and a subsequent linear transformation layer that maps the image embedding space to the task embedding space.
By disentangling physical size from image contents, we can flexibly perceive any size of CT volumes with physical size awareness without resampling CT volumes to have a consistent image tensor across different inputs.

Empirically, we predefined four scales of embedding layers; i.e., the volume size of $16 \times 448 \times 320$ with the path size of $4 \times 16 \times 16$, the volume size of $128 \times 448 \times 320$ with the patch size of $16 \times 16 \times 16$, the volume size of $128 \times 192 \times 224$ with the patch size of $16 \times 16 \times 16$, and the volume size of $128 \times 320 \times 448$ with the patch size of $16 \times 16 \times 16$, to encode lung nodule, heart, lung cancer, and other chest abnormalities respectively. It is worth noting that any prior attention to sub-volumes can be further applied within each scale by adding the attention masks to all self-attention layers as what is done in NLP models~\cite{liu2019roberta}. We used the bounding boxes of lungs to make the model attend to lungs only in predicting and characterizing lung nodules. 
For M3FM-Base, M3FM-Large, and M3FM-Huge, the numbers of Transformer layers are 12, 24, and 32, and the sizes of image token embeddings are 768, 1,024, and 1,280, respectively.

\subsubsection{Text Transformer}

Any decent language model can be utilized as the text Transformer in M3FM.
Here the text Transformer was implemented as a Byte-level Byte-Pair-Encoding (BBPE) tokenizer~\cite{liu2019roberta}, a text encoder consisting of the original Transformer layers~\cite{vaswani2017attention}, and a linear transformation layer, as shown in Figure~\ref{fig:method}~(d).
On one hand, the text encoder encodes patient-specific clinical information, such as demographics, smoking history, disease history, cancer history, and other clinical data, which are free text; \textit{e.g.}, ``The patient is 56.0 years old. Gender is Female. Ethnicity is neither Hispanic nor Latino. Height is 60.0 inches. Weight is 105.0 pounds. Education is associate degree/ some college. Former smoker. Smoking duration is 38.0 pack years. Smoking intensity is 20.0 cigarettes per day. 2.0 years since quit smoking. The patient had asthma (childhood) diagnosed at 7.0 years old. The patient had hypertension diagnosed at 53.0 years old. The patient had pneumonia diagnosed at 50.0 years old. Patient's brother(s) (including half-brothers) have lung cancer."
On the other hand, the text encoder encodes free-text task instructions/questions, which are used as the input of the task encoder to extract task-specific embedding features from the multimodal data;
\textit{e.g.}, ``Is there any significant cardiovascular abnormality?'' and ``Predict the lung cancer risk over 6 years.''
This approach allows for embedding any combination of clinical information through free-text prompting, regardless of order. The control signals for specific tasks are then extracted from the text prompts by the same text encoder.
Again, the linear transform maps the text embedding space to the unified task embedding space.
For all our M3FMs, the number of Transformer layers is 12, and the size of text token embeddings is 768.

\subsubsection{Task Encoder}

Figure~\ref{fig:method}~(e) illustrates the task encoder, designed to extract task-specific embedding features from the multimodal token embeddings, given the special $<TASK>$ token embedding.
The task encoder was implemented with multiple Transformer layers and regards all tokens as a single input sequence.
Note that only the special $<TASK>$ token is forwarded to the task encoder and the rest of the question tokens are ignored, as we empirically found that other tokens are not useful in practice.
The special token embedding from the final Transformer layer serves as the task-specific embedding feature that integrates all multimodal data.
For M3FM-Base, M3FM-Large, and M3FM-Huge, the number of Transformer layers is 4 in every case, and the sizes of task token embeddings are 768, 1,024, and 1,280, respectively.

\subsection{Predictors}

The Predictors map task-specific embedding features to answers.
In this study, we found that the task-specific predictor can be automatically and perfectly selected through the Task ID Predictor, which takes the $<TASK>$ embedding corresponding to the question text.
We implemented all Predictors including the Task ID Predictor as a two-layer MLP.
Different tasks may have different Predictors or shared Predictors for the same output dimension, such as "Yes" or "No" answers.
Similar to language models that regard text generation as the token classification over a vocabulary problem, we formulate our answer prediction as a classification problem over the predefined answer candidates, as summarized in Table~\ref{table_mqa}, except for six-year lung cancer risk prediction.
Following \cite{mikhael2023sybil}, we formulate lung cancer risk prediction as a hazard regression problem.
It is worth mentioning that more types of prediction tasks, such as image segmentation and object detection, can be incorporated into M3FM by adding the corresponding lightweight task-specific predictors as demonstrated in our previous study~\cite{niu2023ct}.

\subsection{Self-supervised Pretraining}

A key step to optimize large models is self-supervised pretraining with large unlabeled datasets.
In this study, we adapted the masked autoencoder method~\cite{he2022masked, feichtenhofer2022masked} to pretrain our CTViT on the constructed OpenM3Chest pretraining dataset.
Figure~\ref{fig:method}~(b) shows the CTViT pretraining architecture, which consists of CTViT, an image decoder, and a multi-scale linear de-tokenizer.
The image encoder introduced in Subsection~\ref{sec_imgencoder} was optimized by predicting masked cubes (85\%) from a small number of visible cubes (15\%).
To reduce the memory overhead, only some selected slices along the longitude direction were predicted while recovering each 3D patch.
We pre-trained CTViT with the defined multi-scale 3D CT volumes and a set of data augmentation operations, including random cropping, rotation, resizing, and perturbed display windowing.
The text Transformer in M3FM was initialized with the off-the-shelf RoBERTa model pre-trained via masked language modeling and then trained end-to-end~\cite{devlin2018bert,liu2019roberta}.

\subsection{Multitask Learning}
After self-supervised pretraining, M3FM can be trained with any combination of different tasks with properly selected multimodal datasets by optimizing multitask loss functions simultaneously.
We used the sigmoid cross-entropy loss function for the CVD mortality risk and lung cancer risk prediction tasks and the softmax cross-entropy loss function for all other tasks.
As the number of tasks increases, there is a significant rise in computational cost.
To address this problem, we design a distributed task-parallel (DTP) training strategy.
TDP assigns each computing device with a single task and a single data loader while the total number of training samples remains fixed across all devices for each task.
Since M3FM is a unified model capable of handling various tasks, despite differences in input and output dimensions, gradients computed across all tasks can be readily accumulated, enabling simultaneous parameter optimization.

\subsection{Transfer Learning}

M3FM is designed for adaptability and generalization, enabling the enhancement of out-of-distribution task performance through transfer learning. This capability extends to new tasks with varying image input dimensions, clinical data types, and output dimensions. To accommodate different image dimensions, the addition of a linear embedding layer suffices. For diverse clinical datasets, we can simply describe involved clinical data in free text to the model, without needing any modification on the M3FM architecture, as shown in Figure~\ref{fig:immu}~(a). Specifically, adjusting to different output dimensions requires only the inclusion of a lightweight predictor. Consequently, M3FM can be easily fine-tuned to enable new tasks by leveraging the pre-trained model parameters.

\subsection{Training Details}
We used the AdamW optimizer~\cite{loshchilov2018decoupled}, cosine decay learning rate schedule~\cite{loshchilov2016sgdr}, weight decay of 0.05, and automatic mixed precision in PyTorch for training all models. 
In pre-training CTViT, the batch size was 192, the learning rate was $3.75\times10^{-4}$, on each GPU a single scale of CT input was randomly selected from the four sizes described in Section~\ref{sec_imgencoder}, the CTViT was trained for 200K iterations with 10K warmup iterations, the decoder depth was 2, the voxel values were normalized within each cube in calculating the MSE loss function, and the CT inputs were randomly scaled by the factor of [0.5, 2], [0.6, 1.4], and [0.6, 1.4] in axial, coronal, and sagittal dimensions, and the physical sizes were recalculated accordingly.
In training task-specific models including the transfer learning, the batch size was 12, the number of training iterations was 30K with 2K warmup iterations, the learning rate was $2 \times 10^{-4}$, and the layer-wise learning rate decay of 0.95 was used.
In multitask training, the total batch size was 972, including 12 samples for each of the 17 tasks and 768 samples for the clinical information retrieval tasks.
All CT inputs had a random HU range perturbation, random rotation degrees, and random padding in training M3FM models, with the corresponding hyperparameters in training for different tasks described in Table S4 in the supplementary, and each clinical data element was randomly included with the probability of 0.8.

\subsection{Hardware Requirement}
All our models are trained on the AiMOS Supercomputer in the Center for Computational Innovation at Rensselaer Polytechnic Institute (\url{https://docs.cci.rpi.edu/clusters/DCS_Supercomputer/}).
For CTViT pretraining and multi-task training, we used 192 NVIDIA Tesla V100 GPUs with 32 GiB of memory each, i.e., 6 GPUs per node $\times$ 32 nodes.
For all single-task training and finetuning, we used 12 NVIDIA Tesla V100 GPUs with 32 GiB of memory each, i.e., 6 GPUs per node $\times$ 2 nodes.

\bmhead{Data Availability}
% The raw data of NLST are available from the NIH National Cancer Institute (\url{https://cdas.cancer.gov/nlst/}). The raw data of MIDRC are available at \url{https://www.midrc.org/}.
% Our curated OpenM3Chest dataset will be publicly available upon the acceptance of the paper.
% Restrictions apply to the availability of anonymized patient data from WFUSM and MGH for this project with institutional permission and are, thus, not publicly available.
The raw data of NLST are available in the NIH/NCI Cancer Data Access System (\url{https://cdas.cancer.gov/nlst/}).  The raw data of MIDRC are available at the MIDRC’s Data Access Portal (\url{https://www.midrc.org/}). Our curated OpenM3Chest dataset will be made publicly available upon the acceptance of the paper. Restrictions apply to the availability of the anonymized patient data from WFUSM and MGH for this project with institutional permission, while these data could be shared with potential users through negotiation with each of the institutes.

\bmhead{Code Availability}
Demo codes and data are available at \url{https://github.com/niuchuangnn/M3FM}, and the integrated database and optimized codes will be made open to the academic community upon the publication of our paper.

% \backmatter

\bmhead{Supplementary information}

The accompanying supplementary file contains four tables to show the predefined sentence format for each clinical data element (Table S1), the labeling strategy for each task (Table S2), the predefined questions for each task (Table S3), and the hyperparameters of data transformations for each task (Table S4).
% If your article has accompanying supplementary file/s please state so here. 

% \begin{table*}[h]
% \caption{caption}
% \label{table_sentence}
% \includegraphics[width=1\textwidth]{sentence.pdf}
% \end{table*}

\bmhead{Acknowledgments}

% Acknowledgments are not compulsory. Where included they should be brief. Grant or contribution numbers may be acknowledged.

The imaging and associated clinical data downloaded from MIDRC (The Medical Imaging and Data Resource Center) and used for research in this study was made possible by the National Institute of Biomedical Imaging and Bioengineering (NIBIB) of the National Institutes of Health under contract 75N92020D00021. The content is solely the responsibility of the authors and does not necessarily represent the official views of the National Institutes of Health.

\bibliography{sn-bibliography}% common bib file

\end{document}